\documentclass[11pt, a4paper]{article}
\usepackage[utf8]{inputenc}
\usepackage{geometry}
\usepackage{biblatex} %Imports biblatex package
\usepackage{hyperref}

\usepackage[utf8]{inputenc}
\usepackage{amsmath}
\usepackage{bm}
\usepackage{graphicx}
\usepackage{csquotes}

\title{Using Thermal Ratchet Mechanism to Achieve Net Motility in Magnetic Microswimmers}
\author{Gouri Patil$^1$, Pranay Mandal$^2$, Ambarish Ghosh$^{1,2}$}
\date{%
    $^1$Department of Physics, Indian Institute of Science, Bangalore-560012, India\\%
    $^2$Centre for Nano Science and Engineering,  Indian Institute of Science, Bangalore-560012, India \\[2ex]%
    ambarish@iisc.ac.in
}

\begin{document}

\maketitle

\begin{abstract}
Thermal ratchets can extract useful work from random fluctuations. This is common in the molecular scale, such as motor proteins, and has also been used to achieve directional transport in microfluidic devices. In this work, we use the ratchet principle to induce net motility in an externally powered magnetic colloid, which otherwise shows reciprocal (back and forth) motion. The experimental system is based on ferromagnetic micro helices driven by oscillating magnetic fields, where the reciprocal symmetry is broken through asymmetric actuation timescales. The swimmers show net motility with an enhanced diffusivity, in  agreement with the numerical calculations. This new class of microscale, magnetically powered, active colloids can provide a promising experimental platform to simulate diverse active matter phenomena in the natural world.      

\end{abstract}

\maketitle

%\tableofcontents
Active matter refers to systems of interacting entities that can convert local energy input into independent displacements and meaningful work\cite{ramaswamy2010mechanics,marchetti2013hydrodynamics}. Examples include a wide range of length scales and phenomena\cite{bechinger2016active}, ranging from molecular motors in intracellular biochemical processes to  bacterial colonies and human migration. There have been significant efforts to develop artificial, motile entities that provide test platforms for studying various active matter phenomena. Of special interest are colloidal particles driven by light\cite{villa2019fuel,kummel2013circular}, electric\cite{bricard2013emergence,yan2016reconfiguring,gangwal2008induced}, magnetic\cite{mandal2013observation} field and chemical\cite{paxton2006chemical} reactions, whose activities could be  externally controlled while ensuring the energy transduction occurs independently and locally at the scale of individual colloids. 

Magnetic powering provides unique advantages in the experimental investigation of non-equilibrium systems, with easy, external control over the activity. Previous studies have included on dynamic self-organization of magnetic colloids under external drive, which forms fascinating active materials through collective interactions\cite{wang2019quantifying,han2020emergence,jin2021collective}. The work reported here is fundamentally different, in which we report a magnetically powered active system, where there is motility is induced at the scale of a single colloid.

The only magnetically powered active swimmer reported \cite{mandal2018magnetic,mandal2013observation,patil2021anomalous,fischer2011magnetically} so far, is based on a versatile experimental platform, comprising magnetic, helical nanostructures suspended in a fluid. The colloidal system is stable and maneuverable in most fluids \cite{pal2018maneuverability,pal2020helical,wu2018swarm,servant2015controlled} and powered by small, spatially homogeneous magnetic fields, which complete control over the various experimental parameters, such as the rheological properties of the surrounding medium\cite{ghosh2018helical} \cite{pal2020helical} and the activity. The hydrodynamic flow generated by these swimmers is similar to many living active systems at low Reynolds numbers\cite{purcell1977life}, such as various species of flagellated bacteria. However, these swimmers demonstrate back and forth motion, implying zero net motility and thus classified as \enquote{reciprocal} swimmers, following the terminology in \cite{purcell1977life}. 

Here, we demonstrate a magnetically powered active particle that shows net displacement over a cycle of the magnetic drive, thus qualifying as a \enquote{non-reciprocal} swimmer. Interestingly, the micron-scale swimmers reported here are rendered motile by rectifying ambient thermal fluctuations, akin to the Brownian motors\cite{reimann2002brownian,rousselet1994directional,astumian1997thermodynamics} commonly encountered at much smaller molecular length scales\cite{astumian1998fluctuation}. While thermal ratchets are ubiquitous in microfluidic manipulation\cite{skaug2018nanofluidic,bogunovic2012particle,verleger2012single}, as far as we know, this is the first demonstration of a colloidal particle achieving self-propulsion through the thermal ratchet mechanism.

\begin{figure}
 \centering
    \includegraphics[width=9cm]{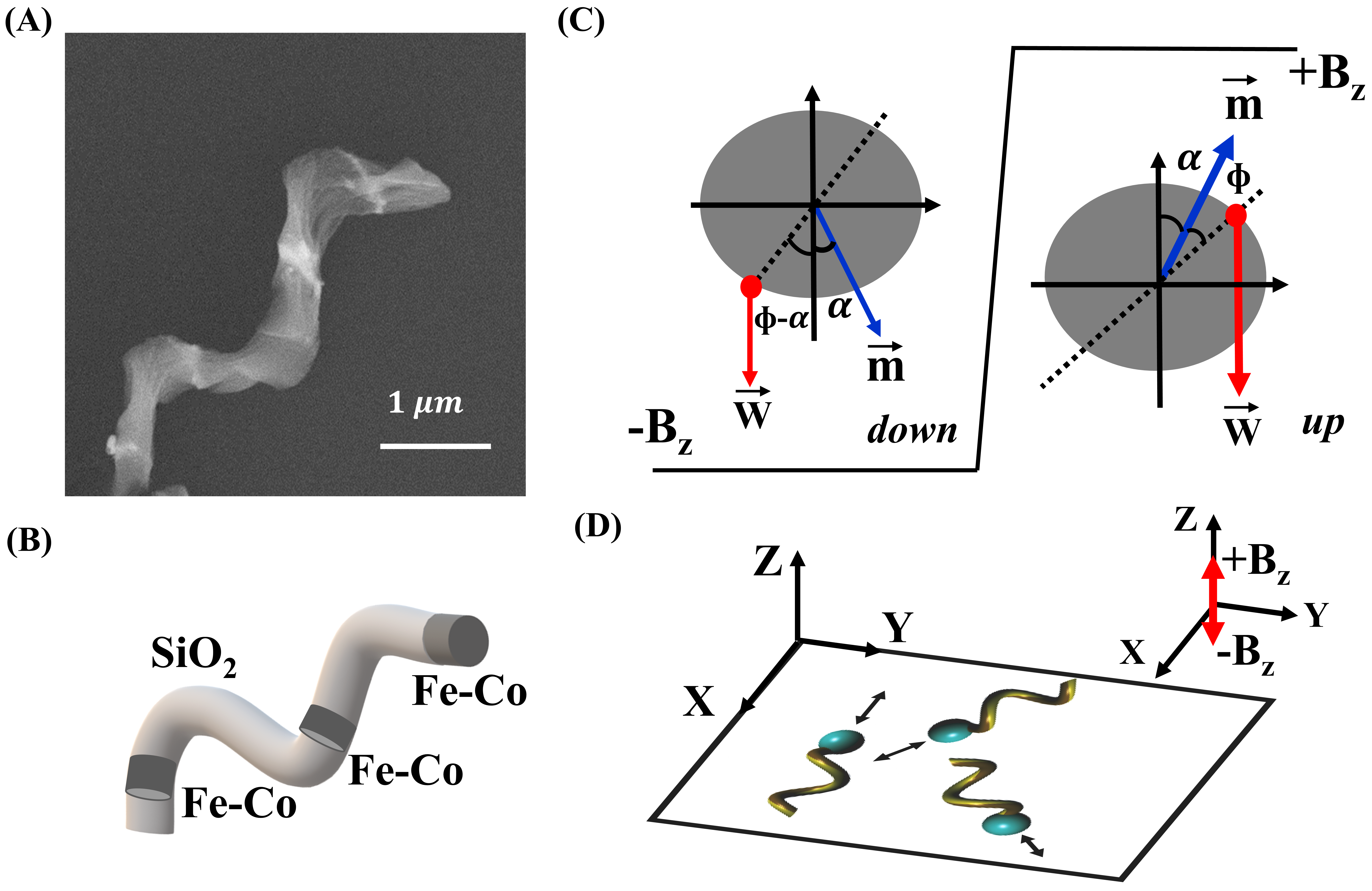}
    \caption{(A) Scanning electron micrograph and (B) schematic of the microhelix showing the material composition. (C) Schematic of the two possible states of the helix in an oscillating magnetic field. The state ('down') corresponds to the field $-B_z$ has the magnetic moment directed downward, while $+B_z$ corresponds to the 'up' state. In the limit of zero thermal noise, the intrinsic mass asymmetry of the helix (shown by $\Vec{W}$) caused the down to up transition through a CCW rotation. We do not show the CW rotation from up to down state, expected when thermal fluctuations are absent. (D) The oscillating field in the z-direction results in a reciprocal motion of the swimmers with no constraints on orientation in the XY plane.}

    \label{fig:1}
\end{figure}

We used a physical vapor deposition technique called GLancing Angle Deposition (GLAD)\cite{hawkeye2007glancing} to fabricate a helically nanostructured film (see supplemental material, section 1) containing magnetic elements embedded in a dielectric (silica) scaffold. The film is sonicated in water to release individual ferromagnetic micro helices (see Fig.\ref{fig:1}A and \ref{fig:1}B for scanning electron micrographs and the schematic) in a microfluidic Hele-Shaw cell. For typical experiments, the helices remained confined in a quasi-2D XY plane, observed under an optical microscope. The helices could be propelled\cite{ghosh2009controlled} like a corkscrew by applying a spatially uniform rotating magnetic field with speeds proportional to the rotational frequency of the field, provided the driving magnetic torque was more significant than the viscous drag. However, we must stress that this mode of actuation does not qualify\cite{mandal2018magnetic} the motion to be self-propelled, as the system is subject to a net external torque. As a consequence, the direction of motion for all helices remains identical, defined by the sense of rotation of the field and the handedness of the helix.  
 
\begin{figure}
    \centering
    \includegraphics[width=8cm]{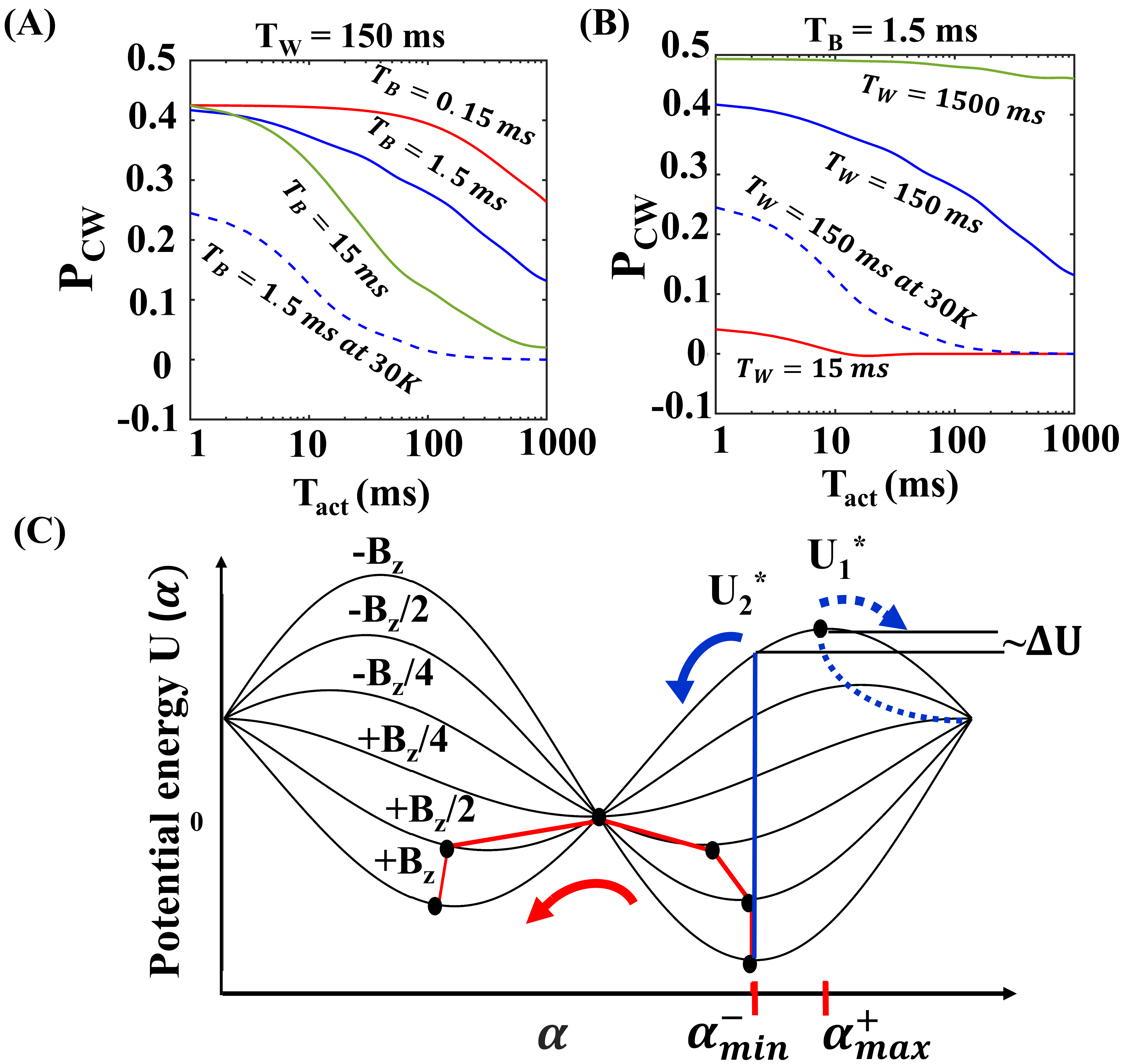}
    \caption{Following the actuation scheme from $-B_z$ to $+B_z$ shown in Fig.\ref{fig:1}C, calculated probability of an anomalous (here, CW) turn as a function of (actuation time) $T_{act}$, for different values of (A) $T_{B}$ and (B) $T_{W}$. Also shown is the effect of thermal noise (temperature) in the same graph. (C) The potential energy of the microhelix as a function of $\alpha$, where red and blue lines show possible paths in configuration space ($\alpha$) taken by the helix. Solid (red and blue) paths correspond to normal (zero thermal fluctuations) turning events, while anomalous events (dotted blue path) can occur at finite temperatures and at low values of $T_{act}$.}
    \label{fig:2}
\end{figure}
The situation changes under a different magnetic drive when the field is oscillated perpendicular\cite{mandal2013observation} to the confining plane, causing the helix to rotate back and forth about its long axis due to an interplay between the magnetic torque and the asymmetric mass distribution of the helices (see Fig.\ref{fig:1}C). These clockwise (CW)-counterclockwise (CCW) rotations resulted in the reciprocal linear displacements corresponding to half of their hydrodynamic pitch ($p_H/2$). As shown in Fig.\ref{fig:1}D, the helix directions were independent, qualifying this system of reciprocal swimmers as a magnetic active matter. In addition, the thermal noise caused imperfection in the CW-CCW sequence, such as CW-CCW-CW-(CW)-CW-CCW. However, the total number of CW and CCW turns remained the same, resulting in an equal number of forward and backward movements when observed over a long time. 

To devise a method to break the reciprocal symmetry of the helical turning sequence and have more CW turns than CCW (or vice versa), we considered the role of thermal noise and actuation timescale $T_{act}$ (see Fig.\ref{fig:1}C) at which the field is switched from between $-B_z$ to $+B_z$. The rotational coordinate $\alpha$ of the helix is governed by two counteracting torques: due to applied magnetic field and mass distribution in the helical structure, as well as intrinsic thermal fluctuations. 
\begin{equation}
    \gamma_{l}\frac{d\alpha}{dt}=-mB_{z}(t)sin(\alpha)+WRsin(\alpha+\phi)+\eta_{r}(t)
\label{eq1}
\end{equation}
Here, $\gamma_{l} \sim10^{-21}kgm^2s^{-1}$ is the rotational friction coefficient about the long axis of the swimmer, $m \sim10^{-16}Am^2$ is the magnetic moment, $WR \sim10^{-20}N$ is the torque due to mass asymmetry, and $\phi$ (31$^{\circ}$ $\pm$ 2.7$^{\circ}$) is the angle between  the direction of magnetic moment and the line of action of weight (see supplemental material, section 2). The uncorrelated noise is given by $\eta_{r}(t)$ with $<\eta_{r}(t)\eta_{r}(t')>=(2k_{B}T/\gamma_{l})\delta(t-t')$, where $k_{B}$ and $T$ correspond to the Boltzmann constant and ambient temperature respectively. The form of $B_z(t)$ is shown in Fig.\ref{fig:1}C with a characteristic actuating time, $T_{act}$. We next re-write equation 1, in terms of $T_W = \gamma_{l}/WR$ and $T_B= \gamma_{l} /mB$; timescales related to magnetic torque and weight respectively:  
\begin{equation}
    \frac{d\alpha}{dt}= \frac{sin(\alpha)cos(\phi)}{T_{W}}+\frac{cos(\alpha)sin(\phi)}{T_{W}}-\frac{sin(\alpha)}{T_{B}}+\eta_{r}(t)    
\end{equation}

We perform a numerical calculation of the sequence ($-B_z \rightarrow +B_z$)  shown in Fig.\ref{fig:1}C, which is expected to be CCW in the absence of thermal fluctuations. We plot the probability of the anomalous CW turn as a function of the actuation timescale in Fig.\ref{fig:2}A and \ref{fig:2}B, at $\phi$ = 31$^{\circ}$, for different values of $T_B$ and $T_W$. The choice of $T_B$, $T_W$ is based on the experimental system (see supplemental material, section 3). Specifically, the curves shown in blue in \ref{fig:2} represent the timescales of our experiment closely, given by $T_B$= 1.5 ms and $T_W$ = 150 ms. All calculations assumed the ambient temperature to be 300 K. As expected, at lower thermal noise with assumed ambient at 30 K, there are fewer anomalous turns (here, CW). Shorter $T_{act}$ implies a greater number of anomalous turns, with a maximum being 0.5, implying an equal number of CW and CCW turns. 

The simulation results and the role of $T_B$, $T_W$ can be understood by considering the energy diagram of the helix as a function of $\alpha$, as shown in Fig.\ref{fig:2}C, and noting its equivalence to Brownian ratchet\cite{astumian1998fluctuation} potentials. The potential energy is given by,
\begin{equation}
    U(\alpha)=-mB_{z}cos(\alpha)+WR(1+cos(\alpha+\phi))
\label{eq3}
\end{equation}
 This is in the form of an additive-multiplicative ratchet potential $U(\alpha)=v(\alpha)+f(t)w(\alpha)$, and as far as we know, the general analytical solutions to this problem are not known\cite{rozenbaum2021exactly}. As shown in Fig.\ref{fig:2}C, higher $T_{act}$ to go from $-B_z$ to $+B_z$ corresponds to the red path in the configuration space, where the system is always at minimum energy configuration. A faster $T_{act}$ implies a non-equilibrium configuration and two possible paths (shown in blue). In the absence of thermal fluctuations, the swimmers take the solid path ($P_{CW}$= 0), while the dashed path can occur due to thermal fluctuations, with a probability that depends exponentially on the magnitude of $\Delta{U}/k_{BT}$, which in turn depends on the experimental system ($\phi$, $T_B$, $T_W$: see supplemental material, section 3). At higher thermal noise or low $T_{act}$, the probabilities of the two paths are equal, implying $P_{CW}$= $P_{CCW}$ = 0.5. 
 
 \begin{figure}
    \centering
    \includegraphics[width=8.5cm]{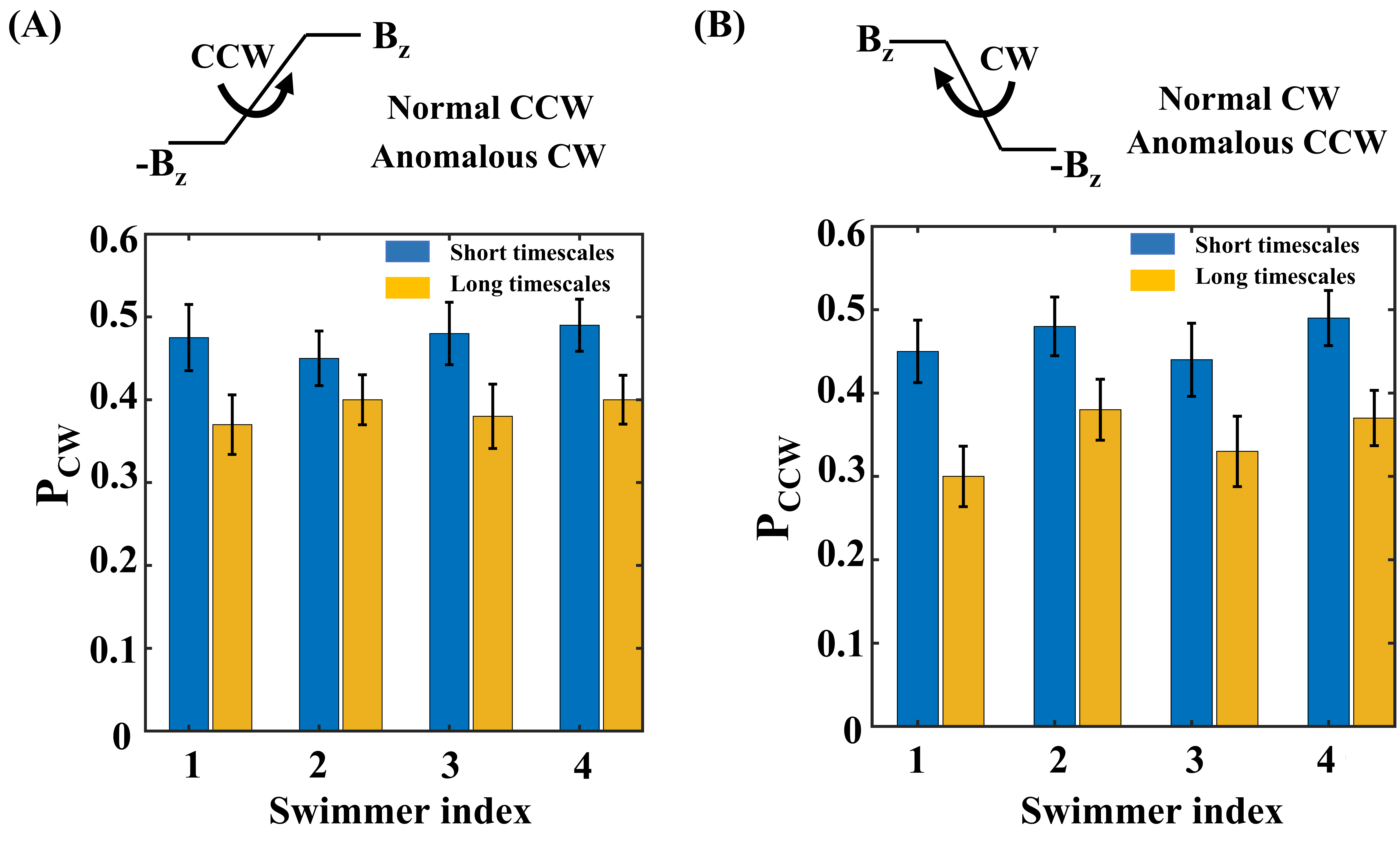}
    \caption{(A) When $B_{z}$ goes from -10 G to +10 G, there are more CW turns compared to CCW turns when the timescale is shorter (blue bars). (B) Conversely, when $B_{z}$ goes from +B to -B, there are more CCW turns compared to CW turns when the timescale is shorter (blue bars)}
    \label{fig:fig3}
\end{figure}

The dependence on $T_{act}$ is further confirmed by experiments, where we observe the rotational dynamics of the helix under magnetic actuation at short (10 – 50 ms) and long (200 – 1000 ms) actuation timescales. Measurements over various swimmers confirm the increased number of anomalous turns at shorter timescales. These experiments required direct visualization of the turning mechanism, so we used a slightly larger helical nanostructure where it was possible to identify the turning direction easily (see supplemental material, section 6). The graph in Fig.\ref{fig:fig3} shows the measured probability of anomalous turnings as a function of the magnitude of actuation time ($T_{act}$), with data from four different swimmers. In Fig.\ref{fig:fig3}(A), when the magnetic field switches from -10 G to +10 G, the probability of anomalous (here CW) turns increases at shorter actuation timescales, which remains consistent across four different swimmers. Similarly, when the field switches from +10 G to - 10 G, the converse (more CCW turns) occurs, as shown in Fig.\ref{fig:fig3}(B).
\begin{figure}
    \centering
    \includegraphics[width=8.5cm]{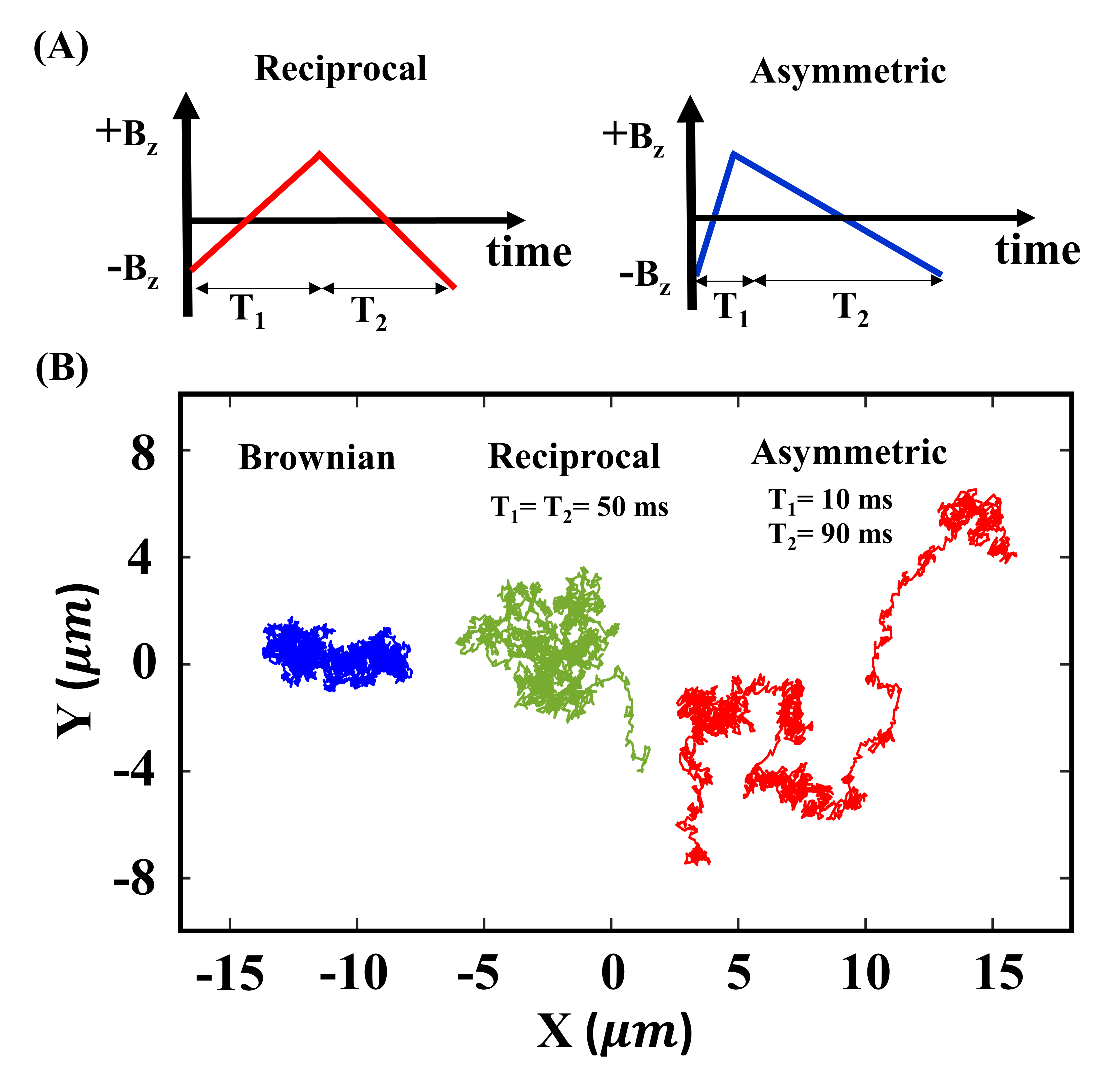}
    \caption{(A) Schematic of the magnetic actuation drive to break reciprocal symmetry. Timescales $T_1 = T_2$ implies equal number of CW turns, while $T_2>T_1$ suggests more anomalous CW turns during the $-B_{z}\rightarrow +B_{z}$ event, and thus having more CW turns overall. (B) Trajectories acquired over 110 seconds for a non-actuated (blue, freely diffusing), reciprocal (green, $T_1$= $T_2$ = 50  ms) and two swimmers under asymmetric drive (red,$T_1$ = 10 ms, $T_2$ = 90 ms).}
\label{fig:4}
\end{figure}
An essential and difficult aspect of the experiments proposed here is to ensure the field in XY plane is negligibly small; otherwise, it results  in alignment and, therefore, the unidirectional motion of the swimmers. This can occur for both dc\cite{mandal2015independent} and ac\cite{leighton1965feynman} fields. We use a tri-axial Helmholtz coil to circumvent this problem and cancel stray magnetic fields in the XY plane. The correction is further confirmed by checking the orientations of the individual swimmers to be independent at all times. As shown in the supplemental material, section 4, we track trajectories of two swimmers under asymmetric drive and subsequently analysed the helix orientations. The orientations were uncorrelated with correlation coefficient = 0.0772. Please note the distance between the swimmers in this experiment was greater than five body lengths, which suggests they were effectively non-interacting.

Having confirmed the dependence of anomalous turns on the actuation timescale, we devise a strategy to break the reciprocal CW-CCW sequence, such as to induce net motion in the helical swimmer. We use an asymmetric magnetic drive, shown schematically in Fig.\ref{fig:4}A. Following the same convention as in Fig.\ref{fig:1}C, we expect a short $T_1$ resulting in more anomalous events, with maximum being $P_{CW}$ = $P_{CCW}$ = 0.5. A larger $T_2$ implies fewer anomalous turns, implying $P_{CW}$ $>$ $P_{CCW}$; thus resulting in a greater number of CW turns over a full cycle of magnetic actuation and breaking the reciprocal symmetry. We show the trajectories (see Fig.\ref{fig:4}B) of a swimmer without magnetic actuation (Brownian motion), reciprocal actuation $(T_1= T_2)$ and asymmetric drive $(T_2:T_1 \neq 1)$, acquired over approximately 110 seconds. The overall increase in the mean squared displacement (asymmetric $>$ reciprocal $>$ Brownian) can be clearly seen. 

\begin{figure}
    \centering
    \includegraphics[width=6cm]{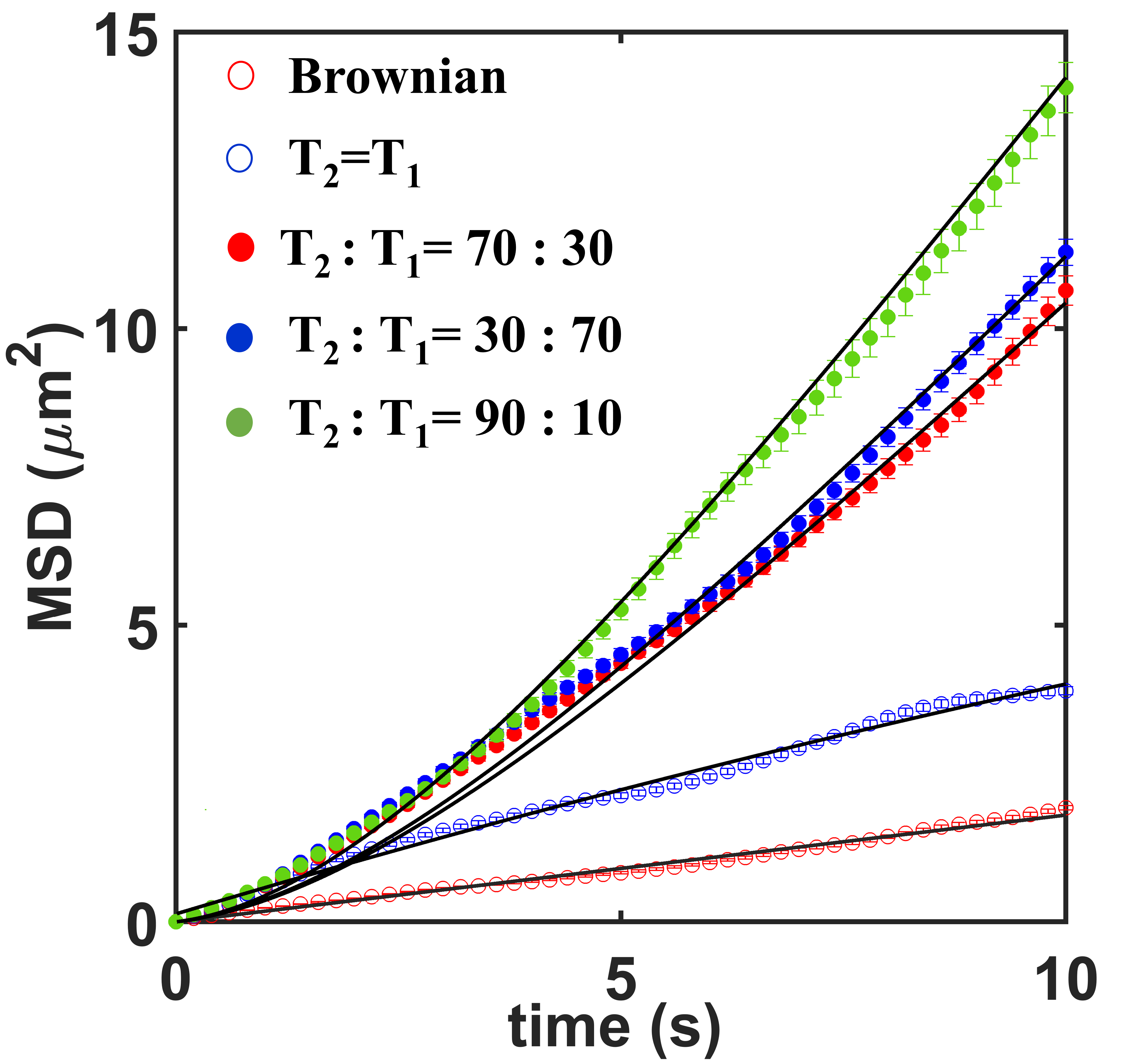}
    \caption{Mean square displacement (MSD) for a swimmer as a function of time, plotted for different $T_2:T_1$, shows greater diffusivities at higher ratios of $T_2:T_1$.  The duration of actuation cycle ($T_1+T_2$) was 100 ms and the curves are fitted with a theoretical fit (see main text). The fit provides effective self-propulsion velocities of 0.540 $\mu m/s$, 0.450 $\mu m/s$, 0.470 $\mu m/s$ for $T_2:T_1$ ratio varying at $9:1$, $7:3$ and $3:7$ respectively}
    \label{fig:5}
\end{figure}

The increase in diffusivity under reciprocal actuation is well understood\cite{mandal2013observation,lauga2011enhanced} and related to the orientation fluctuations of the swimmer in the XY plane. However, the diffusivity increased further as the drive timescales were made asymmetric. This occurred due to an enhanced number of anomalous turns during one cycle of magnetic actuation, resulting in a non-reciprocal stroke pattern, and thus forming a swimmer with net motility. In Fig.\ref{fig:5}, we show the mean square displacement for a swimmer (more data for other swimmers available in the supplemental material, section 5) under different levels of actuation asymmetry (given by $T_2:T_1$) and observe the overall diffusivity to indeed increase with higher $T_2:T_1$ values. This easily follows from a larger number of anomalous turns as $T_2:T_1$ increases, as per the numerical simulations shown in Fig.\ref{fig:2}A and \ref{fig:2}B . The results for $T_2:T_1$ = 70 : 30 are similar to $T_2:T_1$ = 30 : 70, which implied comparable number of anomalous turns of opposite handedness (CW or CCW) for the two cases.

The linear fit of the MSD with time for Brownian and reciprocal swimmers yield $D_{k_{B}T}=5\times10^{-14}m^2s^{-1}$ and $D_{reci}=1.2\times10^{-13}m^2s^{-1}$. The increase and oscillatory dependence of the diffusivity on time for reciprocal actuation was in complete agreement with previous report\cite{mandal2013observation}. The new and interesting aspect of the swimmer under asymmetric actuation was further rise in diffusivity, for which we fit the MSD (for $T_1$:$T_2$ = 9) as per the theoretical description\cite{howse2007self,franke1990galvanotaxis} for self propelled particles, where $MSD (\tau) = [4D_T + 2v^2 \tau_r]\tau + 2v^2 \tau_{r}^{2}[e^{-\tau/\tau_{r}}-1]$, where $D_{kT}$ and $\tau_r = 3$ seconds corresponds to the Brownian translational and orientational (XY plane) diffusivities of the swimmer. We estimate an effective self propulsion velocity around 0.5 $\mu m/s$, which could be increased further by manipulating $T_B$ to achieve higher number of anomalous turns.

In summary, we have realized a new class of active swimmers powered by external magnetic fields. Unlike other synthetic microscale swimmers reported before, the motility of these swimmers is derived from thermal fluctuations using the Brownian ratchet mechanism, which suggests similar strategies may be adopted to develop artificial swimmers at smaller length scales. These self-propelled helical swimmers can be dispersed in various media, including rheologically complex\cite{ghosh2018helical} and heterogeneous environments\cite{dasgupta2020nanomotors}, and their motility can be controlled by engineering the magnetic drive. It may be possible to add multiple functionalities to these swimmers, including optofluidic manipulation\cite{ghosh2018mobile,patil2021magnetic}. We believe this experimental system is highly suitable for studying and simulating the various phenomenological models for the wet active matter. It may be particularly interesting to study the interplay of hydrodynamic interactions and noise-induced motility in shaping their collective behavior. Also, this system is known to be bio-compatible and maneuverable in various biological media\cite{pal2018maneuverability,pal2020helical,dasgupta2020nanomotors,venugopalan2014conformal,wu2018swarm}, implying futuristic  biomedical applications emerging from self-propelled, theragnostic\cite{venugopalan2018single,vasantha2021theragnostic,nelson2010microrobots,qiu2022magnetic} entities deployed in living systems. 

The authors would like to thank Ajay Ajith for his help with the experiments involving cancellation of stray magnetic fields. We gratefully acknowledge DBT, SERB and MeiTY for funding support, and the usage of the facilities in the Micro and Nano Characterization Facility and National Nano Fabrication Centre (CeNSE) at IISc.

\printbibliography

\end{document}

\maketitle
\section*{Content}

\section*{(1) Fabrication and actuation mechanism}

\section*{(2) Estimating the weight and angle $\phi$}

\section*{(3) Calculations regarding the ratchet potential}

\section*{(4) Low correlation between the swimmers' orientation}

\section*{(5) Diffusivity increase with asymmetric actuation: more examples}

\section*{(6) Movies available}

\section{Fabrication and actuation mechanism}
\begin{figure}
    \centering
    \includegraphics[width=12cm,height=7cm]{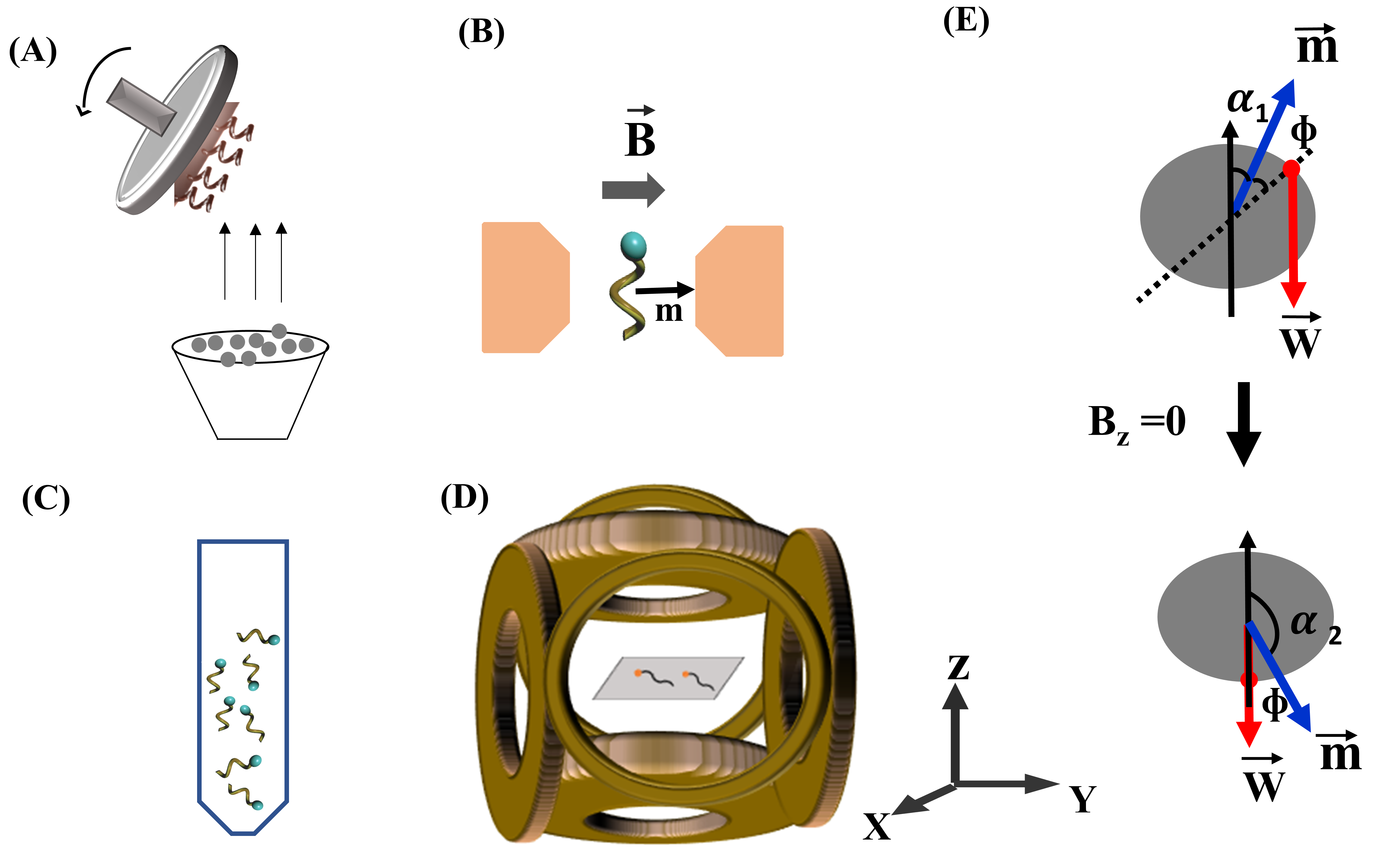}
    \caption{(A) The schematic of a GLAD set up. (B) The swimmers magnetized along the short axis using a permanent magnet. (C) The swimmers dispersed in the deionised water. (D) The Helmholtz coil along with the microfuidic device placed at its centre. (E) The schematic depicting line of action of weight and magnetic moment.}
    \label{fig:coil}
\end{figure}
The magnetic swimmers are fabricated using the Glancing Angle Deposition (GLAD) method\cite{hawkeye2007glancing}. This is a unique technique to grow wafer-scale micron-sized columnar structures by a physical vapour deposition method, where a substrate is held at an extreme angle (Fig. \ref{fig:coil} A) with respect to the incoming evaporated plume. The swimmers used for all the experiments in this study are primarily made of silica and small amounts of magnetic material. The helical swimmers are  3 $\mu m$ in length and have a geometric pitch around 1700 $nm$. The iron-cobalt in the ratio of 1:1 is deposited in between the helical structure along with silver which acts as an adhesive material between the metal and silica.\cite{patil2021magnetic} The metallic part is distributed across the helix in different positions as shown in the Fig. 1(B) of the main text as a schematic. 

The helical swimmers on the silicon wafer are magnetised along their short axes using a permanent magnet as shown in Fig. \ref{fig:coil}(B). The wafer is placed in a tube of de-ionized (DI) water and the tube is dipped in a sonication bath to release the structures into the DI water (Fig. \ref{fig:coil} C). A small volume of this solution is placed in a microfluidic chamber made of piranha-cleaned glass slides. The chamber is sealed to avoid any flow. To obtain a magnetic field, triaxial Hemholtz coil (Fig. \ref{fig:coil}(D)) is used, a device for producing a region of nearly uniform magnetic field. According to our convention, the triaxial coil is mounted on the microscope such that the $X$ and $Y$ coils generate field in the plane of imaging which is also the plane of the sample and $Z$ coil is in a plane perpendicular to the imaging plane. A signal generated by Labview programming is fed into the NI-DAQ card which relays the information to the amplifiers and then the coils.

\section{Estimating the weight and angle $\phi$}
In Fig. (\ref{fig:coil}E), the schematic shows the propeller with a magnetic moment vector $m$ and the weight $W$. Due to the slight asymmetric weight distribution in the helical structure, the direction of magnetic moment and the line of action of weight do not coincide and the angle between them is denoted by $\phi$. We re-write the equation from the main text:
\begin{equation}
    \gamma_{l}\frac{d\alpha}{dt}=-mB_{z}(t)sin(\alpha)+WRsin(\alpha+\phi)+\eta_{r}(t)
\label{eq1}
\end{equation}
Here, $\gamma_{l} (\sim10^{-21}kgm^2s^{-1})$ is the rotational friction coefficient about the long axis of the swimmer, $m (\sim10^{-16}Am^2)$ is the magnetic moment, $WR$ is the typical torque due to mass asymmetry, and $\phi$ is the angle between the direction of magnetic moment and the line of action of weight. The noise term is considered to be $\eta_{r}(t)$ with $<\eta_{r}(t)\eta_{r}(t')>=(2k_{B}T/\gamma_{l})\delta(t-t')$ where $k_{B}$ and $T$ are the Boltzmann constant and the ambient temperature. We can estimate the magnetic moment by finding the step-put frequency at which the viscous drag is equal to the applied magnetic torque\cite{ghosh2012dynamical,zhang2009artificial}. However, we have two unknowns here ($WR$ and $\phi$) which need to be extracted from our experimental data, as outlined below.  

Under the application of a vertical magnetic field ($+B_z$ or $-B_z$), the helix orientation is determined by two counteracting torques due to applied magnetic field and weight asymmetry. When the field is turned off, the helix re-orients to a new equilibrium configuration governed by the weight alone. The time taken for this rotation, given by $T_1$ and $T_2$ for the two magnetic field directions, can be estimated by integrating the equation \ref{eq1}. The two cases correspond to $\alpha$ changing from $\alpha_1$ to $\alpha_2$ and second, when $\alpha$ changes from $\alpha_2$ to $\alpha_1$. 
\begin{equation}
\begin{split}
    tan\frac{\alpha_{1}+\phi}{2}=exp(-\frac{WRT_{1}}{\gamma_{l}}) \\
    tan\frac{\alpha_{2}+\phi}{2}=exp(-\frac{WRT_{2}}{\gamma_{l}})
\end{split}
\label{eq2}
\end{equation}
With,
\begin{equation}
\begin{split}
    cot\alpha_{1}=\frac{(\frac{mB}{WR}-cos\phi)}{sin\phi} \\
    cot\alpha_{2}=\frac{(\frac{-mB}{WR}-cos\phi)}{sin\phi}
\end{split}
\label{eq3}
\end{equation}

Subsequently, we measure the times $T_{1}$ and $T_{2}$ experimentally. A magnetic field $B_z = 50 G$ is given in the Z-direction and the field is switched off. This will allow the swimmer to turn solely due to the weight and we measure the time taken $T_{1}$) from the recorded videos. Similarly, $-B_{z}$ is given and switched off to obtain $T_{2}$. This is repeated for ten different swimmers and these experimental times are then used get the information of $WR$ and $\phi$ using eqn \ref{eq2} and eqn \ref{eq3}. We obtained the WR=$(9.5\pm 1.5)\times10^{-21} N$ and $\phi=31^{\circ}\pm2.7^{\circ}$.

\section{Calculations regarding the ratchet potential}
The dynamics of a propeller in the absence of noise is given by 
\begin{equation}
    \gamma_{l}\frac{d\alpha}{dt}=-mB_{z}(t)sin(\alpha)+WRsin(\alpha+\phi)
\end{equation}

Further modifying the equation,
\begin{equation}
\begin{split}
    \frac{d\alpha}{dt} & =\frac{[WRsin(\alpha)cos(\phi)+Wcos(\alpha)sin(\phi)-mB_{z}(t)sin(\alpha)]}{\gamma_{l}} \\
     & =sin(\alpha)\left[\frac{WRcos(\phi)}{\gamma_{l}}-\frac{mB_{0}}{\gamma_{l}}\frac{B_{z}(t)}{B_{0}}\right]+cos(\alpha)\left[\frac{WRsin(\phi)}{\gamma_{l}}\right] 
\end{split}
\end{equation}

When the magnetic field switches rapidly, the helix can follow either of the two pathways and depends exponentially on the magnitude of $\Delta U/k_{B}T$ as shown in Fig. 2C of the main text. The value of $\Delta U$ is given by calculating the energy values at $\alpha^{-}_{min}$ and $\alpha^{+}_{max}$.
\begin{figure}
    \centering
    \includegraphics[width=12cm]{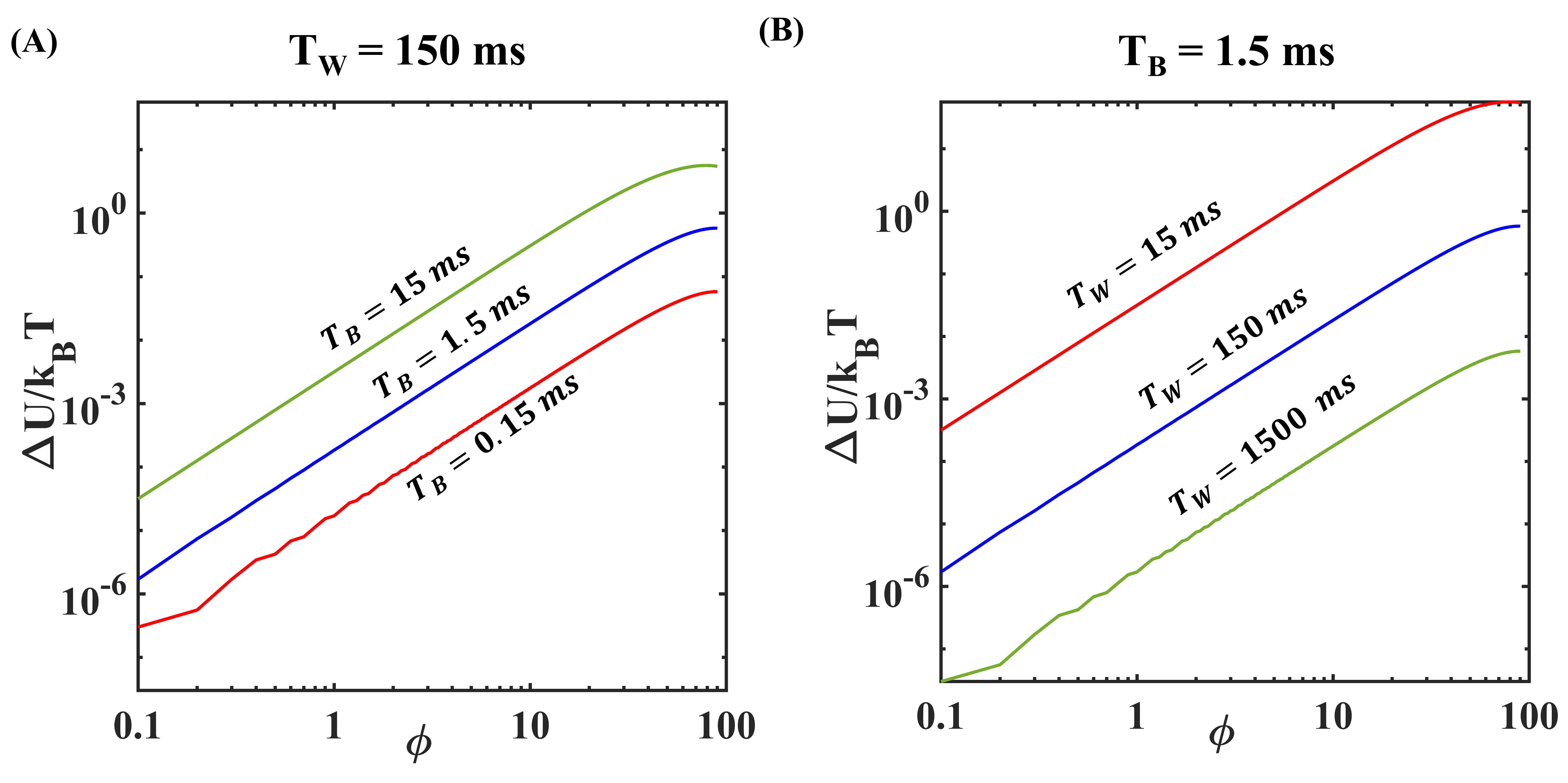}
    \caption{The dependence of $\Delta U$ on $\phi$ is calculated at different $T_W$ and $T_B$.}
    \label{fig:energy}
\end{figure}
Solving the equation $\delta U/\delta\alpha=0 $, we arrive at the values of $\alpha^{-}_{min}$ and $\alpha^{+}_{max}$. 
\begin{equation}
   \alpha^{-}_{min}=\pi-tan^{-1}\left[\frac{T_B sin(\phi)}{T_W}+cos(\phi)\right] 
\end{equation} 
\begin{equation}
   \alpha^{+}_{max}=\pi+tan^{-1}\left[\frac{T_B sin(\phi)}{T_W}-cos(\phi)\right] 
\end{equation} 

As evident from the expressions, $\Delta U= | U(\alpha^{-}_{min}) - U(\alpha^{+}_{max}) |$ is dependent on $T_W$, $T_B$ and $\phi$. This is shown in the Fig.\ref{fig:energy} assuming $T_B$=1.5 ms and $T_W$=150 ms, which were around the experimental values, for the data shown as blue in Fig. \ref{fig:energy}. The probability of anomalous turns is proportional to $exp(-\Delta U/k_{B}T)$. 

\section{Low correlation between the swimmers' orientation}
As mentioned in the main text, it must be ensured that there is no magnetic field in the XY plane to avoid the swimmers aligning in a fixed direction. Though a small field is present in the XY plane, we cancel both the DC and AC components of the field using linear feedback techniques.
 
\begin{figure}[!htb]
    \centering
    \includegraphics[width=10 cm]{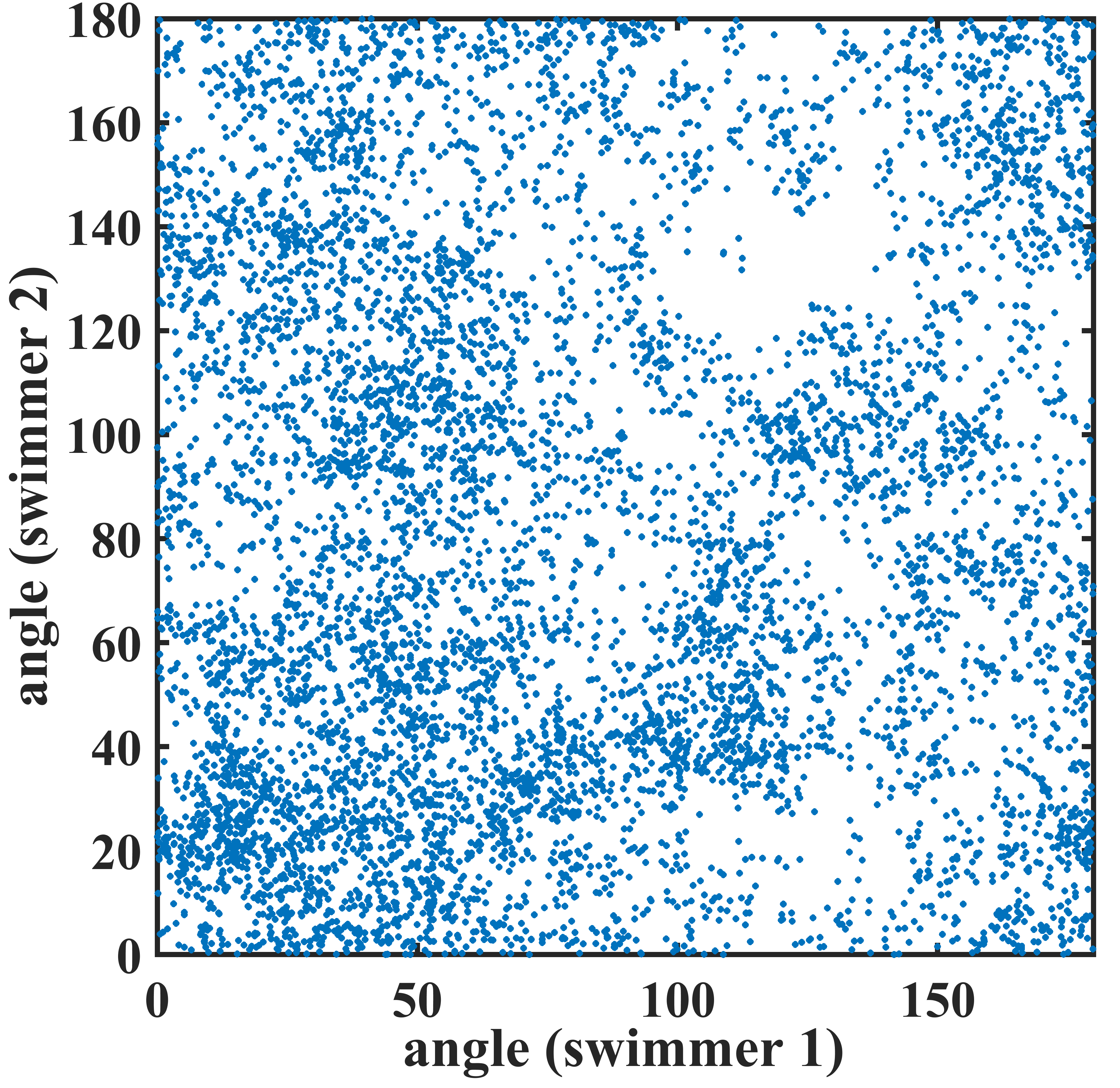}
    \caption{Orientation of two swimmers plotted against each other indicates clearly that there is no correlation between the two.}
    \label{fig:angle}
\end{figure}
Fig. \ref{fig:angle} shows a plot for two swimmers simultaneously recorded while they were being actuated using an asymmetric magnetic field as discussed in the main text. The graph shows that the orientations of both the swimmers are uncorrelated; hence  making them independent in XY plane. The coefficient of correlation was calculated to be about 0.0772 which is quite low. 

\section{Diffusivity increase with asymmetric actuation: more examples}
\begin{figure}[!htb]
    \centering
    \includegraphics[width=12 cm]{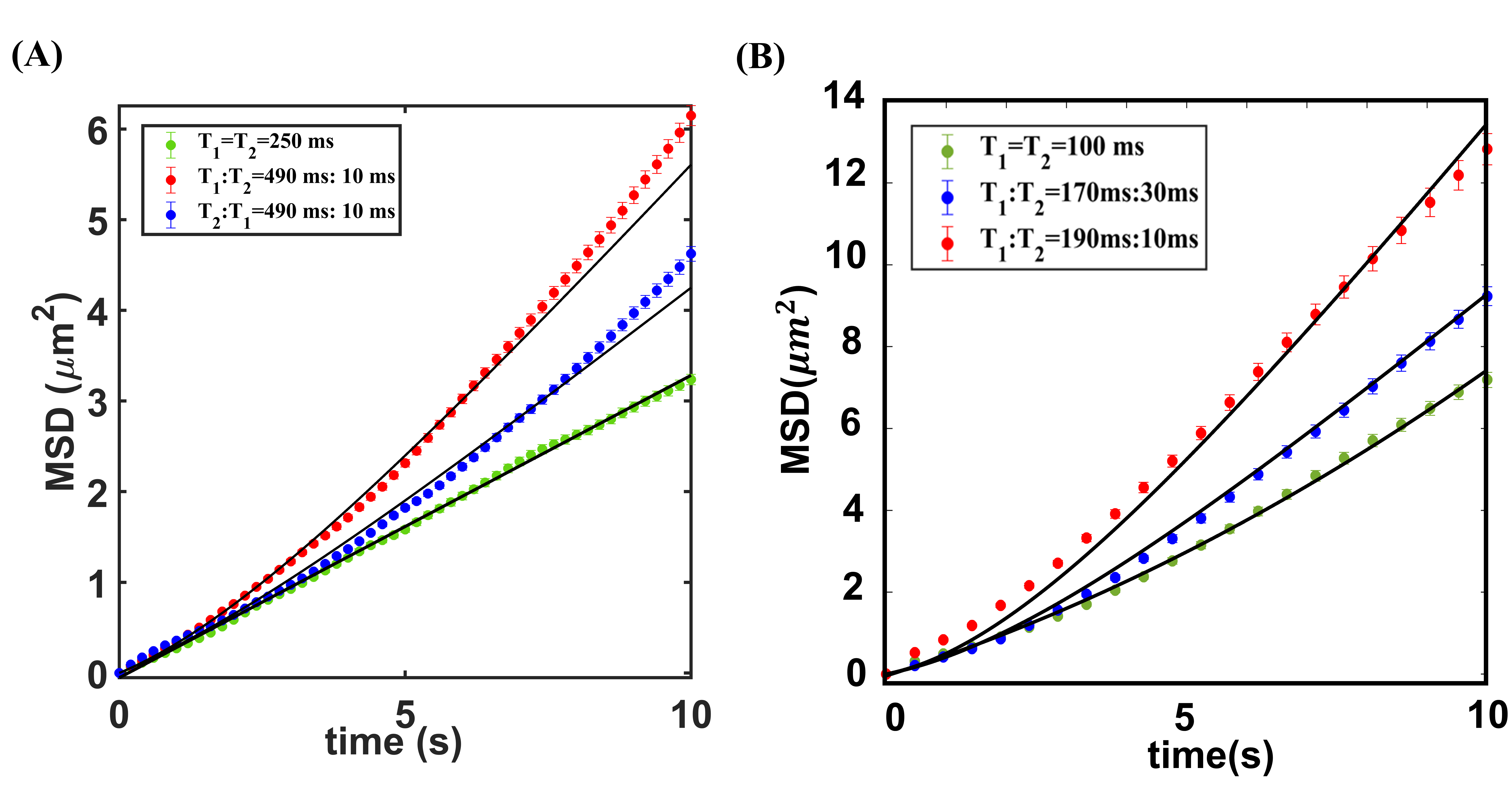}
    \caption{Mean squared displacement for a swimmer with varying ratios of $T_1$ and $T_2$.}
    \label{fig:msd}
\end{figure}
The graph in Fig.\ref{fig:msd} shows that the diffusion increases with increase in asymmetric time scales. We use the following equation similar to the main text to calculate the effective diffusivities of the swimmers driven by asymmetric magnetic drive: $MSD (\tau) = [4D_T + 2v^2 \tau_r]\tau + 2v^2 \tau_{r}^{2}[e^{-\tau/\tau_{r}}-1]$, where $D_{kT}$ and $\tau_r = 3$ seconds corresponds to the Brownian translational and orientational (XY plane) diffusivities of the swimmer.The swimmer with $T_1:T_{2}=490:10 ms$ and $T_1:T_{2}=10:490 ms$ has a much higher diffusion (0.218 $\mu m^2/s$ and 0.178 $\mu m^2/s$) with velocities 0.3 $\mu m/s$ and 0.25 $\mu m/s$ respectively compared to $T_1:T_{2}=250:250 ms$ (0.085 $mu m^2/s$). Similarly in Fig.\ref{fig:msd}(B), another swimmer with $T_1:T_{2}=190:10 ms$ and $T_1:T_{2}=170:30 ms$ has a higher diffusivity (0.445 $\mu m^2/s$ and 0.298 $\mu m^2/s$) with velocities 0.5 $\mu m/s$ and 0.390 $\mu m/s$ respectively compared to $T_1:T_{2}=100:100 ms$ with diffusivity 0.187 $\mu m^2/s$. This indicates that net motility can be achieved using an asymmetric magnetic drive.

\section{Movies available}
\subsection*{S1} 
The movie is recorded at 40 fps and slowed down to 20 fps. Here the swimmer experiences a magnetic field of 10 G and $T_1 = 500 ms$ and $T_2 = 50 ms$. The video shows how the swimmer makes a CW (clockwise) and CCW (counterclockwise) turns. To detect the number of CW and CCW turns, a slightly larger swimmer is used as already mentioned in the main text. With further image processing, we detect and count the turnings. 

\subsection*{S2}
The movie is recorded at 50 fps and has been sped to 200 fps. Here the same swimmer is performing reciprocal actuation with $T_1 = T_2 = 50 ms$ and non-reciprocal actuation with $T_1 = 90 ms$ and $T_2 = 10 ms$.

\printbibliography